\documentclass[a4paper,12pt]{article}
\usepackage{a4wide}
\usepackage{times}%
\usepackage {nopageno, 
multicol,times,here,amsmath,amssymb}
\usepackage[utf8]{inputenc}
\usepackage[]{caption2}
\usepackage{graphics}
\usepackage[demo]{graphicx}
\usepackage{hyperref}
\usepackage{lipsum}
\usepackage{wrapfig, blindtext}
\usepackage{comment}
\usepackage{tabularx}
\usepackage{float}
\usepackage{xcolor}
\usepackage{latexsym,amsbsy,amsmath,bm}
\usepackage{microtype}
\mathchardef\mhyphen="2D
\def\vec#1{\bm{#1}}

\def\ll#1#2{\tilde{\lambda}_{#1}.\tilde{\lambda}_{#2}}

\usepackage[backend=biber,style=numeric,maxnames=5]{biblatex}
\begin{document}
\title{
{\bf Tetraquarks: relativistic corrections\\ and other issues}
\vspace{-6pt}}
\author{Jean-Marc Richard\\
{\small \sl Institut de Physique des 2 Infinis de Lyon,
Universit\'e de Lyon, CNRS-IN2P3 \& UCBL}\\[-2pt]
{\small \sl 4 rue Enrico Fermi, Villeurbanne, France}\\[4pt]
Alfredo Valcarce\\
{\small \sl Departamento de F\'\i sica Fundamental e IUFFyM}\\[-2pt] 
{\small \sl Universidad de Salamanca, E-37008 Salamanca, Spain}
\\[4pt]
Javier Vijande\\
{\small \sl Unidad Mixta de Investigaci\'on en Radiof\'\i sica e Instrumentaci\'on Nuclear en Medicina (IRIMED)}\\[-2pt]
{\small \sl Instituto de Investigaci\'on Sanitaria La Fe (IIS-La Fe)}\\[-2pt]
{\small \sl Universitat de Valencia (UV) and IFIC (UV-CSIC), Valencia, Spain}}
\date{\today}
\maketitle

%
\begin{abstract} 
We discuss the effect of relativistic kinematics on the  binding energy of multiquark states. For a given potential, the use of relativistic kinematics lowers the  energy by a larger amount for the threshold made of two mesons than for a tetraquark, so that its binding is weakened. Some other issues associated with exotic hadrons are also briefly discussed.   
\end{abstract}
%
\section{Introduction}
The recent discovery of the $T_{cc}^+$ tetraquark \cite{LHCb:2021vvq,LHCb:2021auc}, with minimal content $cc\bar u\bar d$, which has been echoed at this Conference, has stimulated some renewed interest in multiquark dynamics. 
 
There is a very rich literature about the $QQ\bar q\bar q$ configurations, starting with \cite{Ader:1981db}. So far, the focus was more on the promising chromomagnetic mechanism giving more attractive strength in specific configurations such as the $H=uuddss$ than in the hadrons constituting their threshold. See, e.g., \cite{Jaffe:2004ph}. In \cite{Ader:1981db,Carlson:1987hh}, and several further studies, it was shown that in a strictly flavor-independent potential (sometimes refered to as a static potential), the system $QQ\bar q\bar q$ with masses $MMmm$ becomes bound if the mass ratio $M/m$ is large enough. For a guide to the bibliography, see, e.g., \cite{Richard:2016eis}. 

In the following, we will also stress the analogy between the quark model in the chromoelectric limit and the pattern of hydrogen-like molecules $M^+M^+m^-m^-$ when the proton-to-electron mass ratio is varied. Actually, the stability of tetraquarks by a pure chromoelectric mechanism is rather elusive, and in actual quark model calculations, the stability of $cc\bar u\bar d$ is due to a cooperative effect of both chromoelectric and chromomagnetic  contributions. See, e.g., \cite{Janc:2004qn,Barnea:2006sd}. 

Interestingly, the stability of some $QQ\bar q\bar q$ is also obtained in lattice calculations (see,e.g., \cite{Bicudo:2019mny,Pflaumer:2021ong} and refs.\ there) and in QCD sum rules
\cite{Navarra:2007yw}. The $T_{cc}^+$ has also been anticipated from the attractive character of the $DD^*$ interaction is some partial waves \cite{Manohar:1992nd,Tornqvist:1991ks,Tornqvist:1993ng,Ericson:1993wy}. This latter approach, sometimes referred to as ``molecular'', is also at work to describe hidden-charm resonances such as $X(3872)$. 

Meanwhile, the quark model has been improved in several ways. The simple variational calculations of the early speculations have been superseded by elaborate hyperspherical expansions \cite{Barnea:2006sd} or methods based on correlated Gaussians \cite{RevModPhys.85.693}. The sophisticated approach of ``real scaling'' gives access to resonances \cite{Meng:2020knc}. See, e.g., \cite{Richard:2018yrm} for an analysis of various aspects of the few-body dynamics within multiquarks. Also, while the potential is usually assumed to be pairwise, with a color dependence corresponding to the exchange of a color octet, alternatives have been studied, where the confining part corresponds to a Steiner tree that generalizes the straight string linking the quark and the antiquark inside a meson, and the  $Y$-shape potential \`a la Fermat-Torricelli joining the three quarks in baryons. 

The four-body problem for tetraquarks is rather involved, as for most other systems. It is thus tempting to attempt some simplifications, which, if justified and successful, shed some light on the structure. 

For instance,  Born-Oppenheimer approach provides an effective $QQ$ potential for $QQ\bar q\bar q$, which indicates whether or not there exists some narrow radial or orbital 
excitation below the dissociation threshold. 

Popular but very controversial is the diquark approximation, in which  baryons and tetraquarks are pictured as quark-diquark and diquark-antidiquark two-body systems, respectively. If this corresponds to solving the few-body problem by steps, first two quarks forming a diquark, it is rarely a good approximation, and it might even induce fake stable multiquarks starting from a Hamiltonian that does not support any. Of course, if the model consists of diquarks at its very beginning, this is different phenomenology that has to be confronted to the data. 

\section{Effect of relativistic kinematics}
\subsection{The case of mesons}
\begin{figure}[H]
 \centering
 \includegraphics[width=.4\columnwidth]{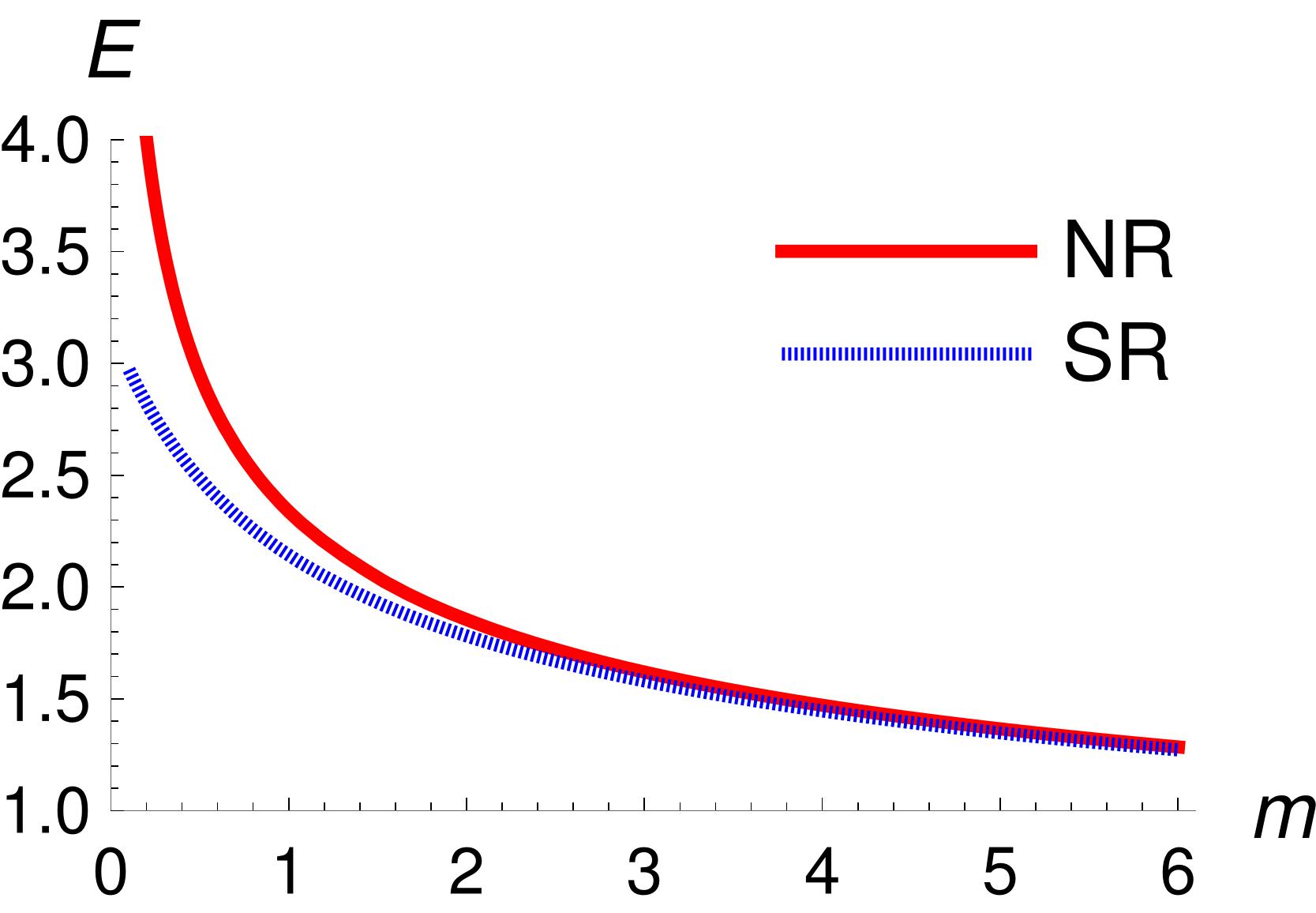}
\caption{Comparison of the non-relativistic and relativistic energies for a $q\bar q$ system as function of the mass $m$ of the quark and the antiquark.}
 \label{fig1}
\end{figure}

Let us consider first a simple model describing a $q\bar q$ meson with the same mass $m$ for the quark and the antiquark. It corresponds to the Hamiltonian
\begin{equation}
 H=K(\vec{p},m)+K(-\vec{p},m)+v(r)~,
\end{equation}
where $v(r)$ is a typical quarkonium potential, e.g., $v(r)=-a/r+b\,r$, and the individual kinetic energy $K$ reads either
\begin{equation}
 K_\text{NR}=\frac{\vec{p}^2}{2\,m}~,\quad\text{or}
 \quad K_\text{SR}=\sqrt{\vec p^2+m^2}-m~.
\end{equation}
for the non-relativistic (NR) and semi-relativisitc (SR) cases, respectively. 
In the pure linear case $a=0$ and $b=1$, the NR and SR energies are compared in Fig.~\ref{fig1}.

Note that in the case of a pure Coulomb interaction ($b=0$), the semi-relativistic Hamiltonian, often referred to as the Herbst Hamiltonian, cannot support too large values for the strength $a$. See, e.g., \cite{Raynal:1993qh}, and refs.\ there.  

The results shown in Fig.~1 have been obtained from a (converged) variational estimate based on a wave function
\begin{equation}
 \Psi=\sum_n \gamma_n\, \exp(-a_n\,\vec r^2/2)~,
\end{equation}
with a suitable minimization of the parameters. For the generalization to baryon and multiquarks, the trial wave function reads
\begin{equation}\label{eq:Gauss}
 \Psi=\sum \gamma_i\, \left[\exp\left(-\sum_{ij}a_{n,ij}\,\vec r_{ij}^2/2\right)+\cdots\right]~,
\end{equation}
where the dots stand for Gaussians deduced by permutations or charge conjugation, in the case of systems with symmetries. This is the method of  correlated Gaussians, which is nowadays widely used in several domains~\cite{2013RvMP...85..693M}. 
\subsection{The case of tetraquarks}
\begin{figure}[H]
 \centering
 \includegraphics[width=.4\columnwidth]{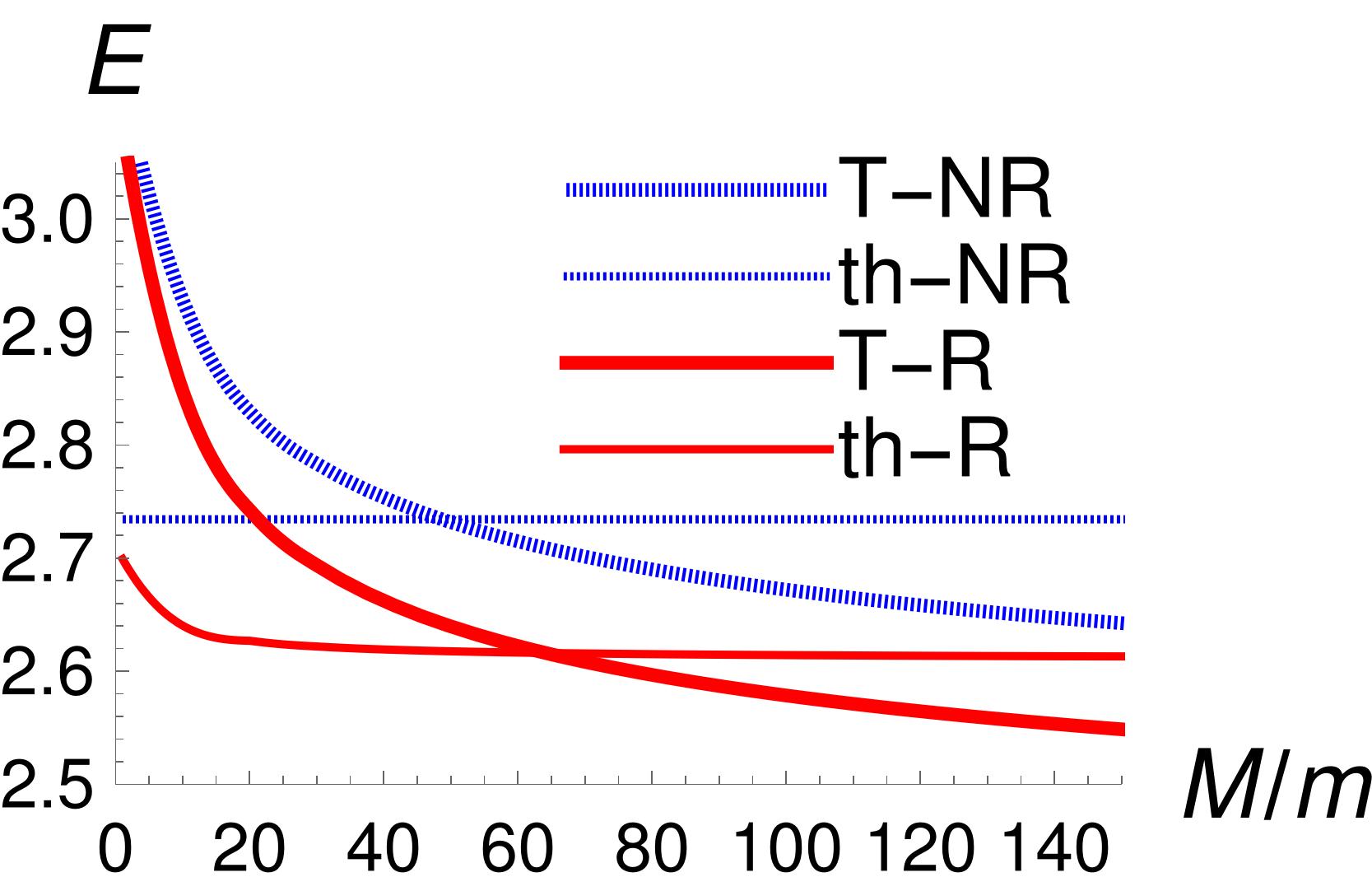}
\caption{Comparison of the NR and SR energies for a $QQ\bar q\bar q$ system as function of the quark-to-antiquark mass ratio $M/m$. The model is that of Eq.~\eqref{eq:H-tetra} with $v(r)=r$. The color part of the wave function is frozen to $\bar 3 3$.}
\label{fig2}
\end{figure}

The calculation can be repeated for baryons and multiquarks. For a tetraquark $QQ\bar q\bar q$ with masses $\{m_i\}=\{M,M,m,m\}$, the Hamiltonian reads
\begin{equation}\label{eq:H-tetra}
H=\sum_i K_i(\vec{p}_i,m_i)-\frac{3}{16}\sum_{i<j} \ll{i}{j} v(r_{ij})~,
\end{equation}
where the $\ll{i}{j}$ operator corresponds to the exchange of a color octet. One can start with a purely linear model, where $v(r)=b\,r$. In the NR, one can rescale so that $a=m=1$. In the SR case, one can still assume $a=1$ without loss of generality. The NR and SR energies are shown in Fig.~\ref{fig2}, as a function of the mass ratio $M/m$, with $M^{-1}+m^{-1}$ fixed, so that the NR threshold remains constant. \\[.1cm]

The main results are in order. All energies are lowered, as a simple consequence of the operator identity $K_\text{SR}\le K_\text{NR}$. However, when bound, the tetraquark energy is less affected than the threshold energy. Thus the binding energy is reduced. And, when the mass ratio $M/m$ is changed, a larger value is required to achieve stability. 

The same pattern is observed for more complicated potentials such as Coulomb-plus-linear, and for the case where the mixing of color $\bar 3 3$ and $6\bar 6$ is taken into account. For more details, and in particular the case where hyperfine effects are included, see \cite{Richard:2021lce}. 
\section{Comparison with atomic physics}\label{sec:atom}
There is an interesting analogy between the hydrogen-like molecules in atomic physics and the tetraquarks in the quark model, at least in the chromoelectric limit. Let us stress a few points. 

\textsl{1)} It is worth noting that the stability is much improved when one goes from the positronium molecule Ps$_2$ ($e^+e^+e^-e^-$) to the hydrogen one, H$_2$ ($ppe^-e^-$). For Ps$_2$, the internal annihilation is disregarded here. In this case, the excess of binding, with respect to the threshold, is about $3.2\,\%$, while for H$_2$, it is increased up to about $17.4\,\%$. The reason has been given by Adamowski et al.\  \cite{1971SSCom...9.2037A,1993PhRvL..71.1332R}. 

The Hamiltonian of H$_2$-like systems is split into a $C$-even part  and a $C$-odd part, say $H=H_0+H_1$, or more explicitly
\begin{equation}
 H=\left[\left(\frac{1}{4\,M}+\frac{1}{4\,m}\right)\sum{\vec{p}_i^2 }+V\right]
 {}+\left[\left(\frac{1}{4\,M}-\frac{1}{4\,m}\right)(\vec p_1^2+\vec p_2^2-\vec p_3^2-\vec p_4^2)\right]~,
\end{equation}
where $V$ denotes the total potential. 
The Hamiltonian $H$ and its even part $H_0$ have the \emph{same} threshold. Moreover, for their ground states, $E(H)\le E(H_0)$ if one uses the variational principle with the solution of $H_0$ as trial wave function. 

Moreover, we note that in this reasoning, the Coulomb character of $V$ hardly matters. The same inequality on the ground state energies is obtained for $V$ being a chromoelectric interaction between quarks and antiquarks, provided it is \emph{flavor-independent}. 

\textsl{2)} In the case of equal masses, the  Ps$_2$ molecule is bound, while $qq\bar q \bar q$ is not bound in  the chromoelectric limit \eqref{eq:H-tetra} of quark models. This can be understood again from the variational principle: if one considers four particles in an attractive potential $v(r)$ with a given cumulated strength $\sum_{i<j} g_{ij}$, the highest ground state energy $E(\{g_{ij}\})$ is obtained in the symmetric case where all $g_{ij}$ are equal, and the less symmetric is the $g_{ij}$ set, the lower the energy. Details are given in \cite{Richard:2018yrm}. 

\textsl{3)} The H$_2$ molecule has never been convincingly described as a diproton linked to a dielectron. 

\textsl{4)} The calculation of four-body bound states is, indeed, rather delicate for systems at the edge between stability and instability. When Wheeler  suggested the existence of Ps$_2$ \cite{1946NYASA..48..219W}, he acknowledged that with a simple wave function
\begin{equation}\label{eq:Wheeler}
 \Psi=\exp\Bigl[-a (r_{12}^2+r_{34}^2)-b\sum_{i\le2\;j\ge3}r_{ij}^2\Bigr]~,
\end{equation}
the best variational energy $E\simeq -0.367$ (in atomic units where $m=\hbar=\alpha=1$) just demonstrates its stability with respect to the dissociation into an isolated electron and a positronium ion $e^+e^+e^-$.  The stability vs.\ two positronium atoms requires 
breaking the symmetries in \eqref{eq:Wheeler} and restoring them as in \eqref{eq:Gauss} by counterterms, but at least four of such combinations of Gaussians are necessary. Thus one admires even better the \textsl{tour de force} by Hylleraas and Ore who demonstrated analytically the stability of Ps$_2$ \cite{PhysRev.71.493}.\footnote{The reference will not be given. Once, a physicist of a great US university criticized \cite{PhysRev.70.90} on the basis that the authors did not remove the center-of-mass motion. This is of course not justified, as the variational wave function of Hylleraas and Ore is translationally invariant, so that the expectation value of the whole Hamiltonian and of the intrinsic Hamiltonian coincide.}\@ For a state of the art on the Ps$_2$ binding energy, see, e.g., \cite{2006PhRvA..74e2502B}. 
\section{Outlook}
A persistent concern by Nathan Isgur was to remove ``naive'' and ``non-relativistic'' from the naive non-relativistic quark model, to understand why it is so successful~\cite{Capstick:1986kw}. We have shown that for multiquarks, the relativistic effects cancel a large fraction of the ones at work in the threshold:  the binding energy, although it is not considerably modified, is slightly reduced. 

As for the ``naive'' aspects, one of them is the assumption, implicit in early quark model calculations, that the interaction is pairwise. Then it was realized that if the linear term of the quarkonium potential is interpreted as the energy of a straight string the quark to the antiquark, its generalization to higher configuration is a kind of Steiner tree linking the color charges with the smallest cumulated length. See, e.g., \cite{Richard:2016eis,Richard:2017vry}.

We insist again on that kinematics is just one facet of relativistic corrections to be applied to the quark model. 

\printbibliography

\end{document}